\newcommand{\gsim}{\raisebox{-0.7ex}{$\stackrel{\textstyle >}{\sim}$ }}
\newcommand{\lsim}{\raisebox{-0.7ex}{$\stackrel{\textstyle <}{\sim}$ }}
\def\thalf{{\textstyle{\frac{1}{2}}}}
\def\twothr{{\textstyle{\frac{2}{3}}}}
\def\fourthr{{\textstyle{\frac{4}{3}}}}

\documentstyle[prd,aps,epsfig,epsf,floats]{revtex}

\begin{document}

\title{Neutrino Propagation In Color Superconducting Quark Matter}

\author {Gregory W. Carter$^1$ and Sanjay Reddy$^2$ }
\address{$^1$Department of Physics and Astronomy, SUNY Stony Brook,
NY 11794-3800.\\
$^2$Institute For Nuclear Theory, University of Washington, Seattle,
WA 98195. }

\date{\today}

\maketitle

\begin{abstract}
We calculate the neutrino mean free path in color superconducting quark matter,
and employ it to study the cooling of matter via neutrino diffusion in the
superconducting phase as compared to a free quark phase.  The cooling process
slows when quark matter undergoes a second order phase transition to a
superconducting phase at the critical temperature $T_c$.  Cooling subsequently
accelerates as the temperature decreases below $T_c$.  This will directly
impact the early evolution of a newly born neutron star, should its core contain
quark matter. Consequently, there may be observable changes in the early
neutrino emission which would provide evidence for superconductivity in hot and
dense matter.
\end{abstract}


\section{Introduction} 

In this article we study heat diffusion via neutrinos in dense, color
superconducting quark matter.  Recent theoretical works
\cite{gap,modern_times,phase_transition,strange_quark,propagators,son}
suggest that quarks form Cooper pairs in medium, a natural
consequence of attractive interactions destabilizing the Fermi
surface.  This would likely affect the early
evolution of neutron stars born through a Type II supernova explosion,
where the central role of neutrino diffusion through a
strongly-interacting medium has long been postulated on theoretical
grounds \cite{spnova}.  

Type II (core collapse) supernovae are triggered by the implosion of 
the inner core of a massive star (M$_{\rm star}$ $\sim 8-20$ M$_\odot$), 
when the core mass is on the order of the Chandrashekar mass (M$_{\rm core}$
$\simeq 1.4 $ M$_\odot$).
During the implosion nearly all ($\sim 99$\%) of the enormous gravitational
binding energy ($\sim 10^{53}$ ergs) gained is stored as internal
energy of the newly born, proto-neutron star (PNS). 
The subsequent evolution of the proto-neutron star is driven by neutrino
diffusion.  Temporal and spectral characteristics of the neutrino
emission depend on the rate at which neutrinos
diffuse through the imploded PNS which, at this early stage, is composed of
hot ($T\sim 20-30$ MeV) and dense ($n_B \sim 2-3~n_0$ where 
$n_0=0.16$ fm$^{-3}$) strongly-interacting matter.

Neutrino emission during this phase is a directly observable
feature of a galactic supernova explosion. 
The few neutrinos detected from SN~1987A indicate that neutrinos of mean energy
$\langle E_\nu\rangle \sim 20$ MeV are emitted on a 
diffusion time scale of about $10-20$ seconds. 
It is reasonable to expect that neutrino mean free path
in the inner, denser regions of the star will strongly influence the
temporal characteristics of a supernova neutrino signal. 
Supernova neutrinos can therefore reveal properties of matter at high 
baryon density, at temperatures substantially lower than those expected in 
relativistic heavy ion collisions.

Although the idea of quark pairing in dense matter is not a new one
\cite{gap}, it has recently seen renewed interest in the context of 
the phase diagram of QCD \cite{modern_times}.  Model calculations, mostly based
on four-quark effective interactions, predict the restoration of spontaneously
broken chiral symmetry through the onset of color superconductivity
at low temperatures \cite{phase_transition}.  For much higher densities color
superconductivity is manifest through perturbative gluon exchange
\cite{son,bbs}, which can be calculated systematically suggests that the
phenomenon is robust.  For densities and temperatures relevant to neutron
stars, quark matter is therefore expected to be superconducting.  Models
generally predict an energy gap of $\Delta \sim 100$ MeV for a typical quark
chemical potential of $\mu_q \sim 400 $ MeV.  As in BCS theory, the gap will
weaken for $T > 0$ and at some $T = T_c$ there is a (second-order) transition
to a ``standard'' quark-gluon plasma.  Thus, when cooling from temperatures
greater than critical, the formation of a such a gap in the fermionic
excitation spectrum in quark matter at high density will influence various
neutron star observables -- if neutron stars contain quark matter in their
cores at early time\footnote{ This remains unclear. In particular, the
existence of a finite electron neutrino chemical potential at very early time
has shown to inhibit the appearance of quark matter. See Ref.~\cite{pipelr} for
a review.}.  Two examples recently investigated are the effects on magnetic
fields
\cite{mit_star} and on the thermal evolution of neutron stars at late time
\cite{cooling}, when interior temperatures evolve from $ T \sim 1$ MeV to $ T
\lsim 1 $ KeV.  In this work we shall consider the cooling of quark matter at
earlier times, when temperatures pass through the critical $T_c \sim\Delta$.
Assuming a simplified scenario, we investigate the thermal evolution of the
inner core of a proto-neutron star as it cools via neutrino diffusion during
its first few seconds.

Our main finding is that the neutrino mean free path in the superconducting
phase has a strong dependence on temperature.  An energy gap tends to increase
the neutrino mean free paths exponentially when $T \ll \Delta $.  However in
the intermediate regime, when $T \sim \Delta$, the temperature dependence is
not exponential.  From this we predict a uniquely uneven cooling process for
simplified PNS matter, marked by a slowdown of the cooling rate when the system
undergoes a second-order phase transition.  This is a consequence of the
specific heat being peaked at $T_c$, a standard characteristic of a phase
transition, rather than a large change in the neutrino mean free path.  Since
the energy gap vanishes at this point, the mean free path is {\it not}
significantly modified in the neighborhood of $T_c$.

In Section II the neutrino mean free path, always denoted $\lambda$ in this
work, is computed in a background of superconducting quarks.  Taking a general,
BCS-type model for the energy gap, we relate $\lambda$ to in-medium quark
polarizations via the differential and total neutrino-quark cross sections.
Explicit formulae are derived for the imaginary parts of the vector and
axial-vector polarizations of a relativistic Fermi system with an energy gap.
After assuming BCS-type mean field behavior of the gap and the specific heat,
in Section III we compute a characteristic time for heat diffusion via neutrino
emission from a simple model of superconducting quark matter.  We then consider
some astrophysical consequences and outline the model's applicability to
proto-neutron stars.  Section IV contains our conclusions.

\section{Neutrino-quark Scattering in a Color Superconductor}

The primary process by which heat escapes a proto-neutron star is neutrino
diffusion, and so a significant consequence of color superconductivity in this
context will be modified neutrino propagation.  While noting that the neutrino
production rate will also differ from that of normal matter, in the diffusive
regime one can see that the dominant critical behavior will be a change in the
{\it inelastic} quark-neutrino cross section, since here neutrino production
rates decouple from the transport equation and depend only on the neutrino mean
free path.  In this section we calculate the differential and total cross
sections and then compute the neutrino mean free path in two-flavor quark
matter.  The magnitude of the superconducting gap, $\Delta$, is taken as
arbitrary within a range of values found in recent literature.  Closely
following BCS theory, we assume the gap to be a constant, and calculate
the response functions and neutrino cross sections
in the weak coupling approximation.

The neutral current coupling between neutrinos and quarks, a four-fermion 
effective interaction for energies $E_\nu \ll M_Z$, is written as 
\begin{equation}
{\mathcal{L}}_W = \frac{G_F}{\sqrt{2}}
~\bar{\nu}\gamma_\mu(1-\gamma_5) \nu \, \bar{q}\gamma^\mu(C_V-
C_A\gamma_5) q \,,
\label{nuq}
\end{equation}
where $G_F=1.166 \times 10^{-5}$ GeV$^{-2}$ is the Fermi weak coupling
constant and $C_V$ and $C_A$ are the flavor-specific 
vector and axial vector coupling constants, respectively. 
The differential neutrino scattering cross section per unit
volume in an infinite and homogeneous system of fermions as
calculated in linear response theory is \cite{HW,rpl}
\begin{eqnarray}
\frac {1}{V} \frac {d^3\sigma}{d^2\Omega_3 dE_3} &=&  -\frac {G_F^2}{32\pi^2}
\frac{E_3}{E_1}~
\frac{\left[1-f_{\nu}(E_3)\right]}{\left[1-\exp{\left(-q_0/T\right)}\right]}
~{\rm Im}~(L^{\alpha\beta}\Pi_{\alpha\beta}) \,,
\label{dcross}
\end{eqnarray}
where $E_{1}$ ($E_3$) is the incoming (outgoing) neutrino energy.  The
factor $ [1-\exp(-q_0/T)]^{-1}$ maintains 
detailed balance and the final state blocking of the outgoing
neutrino is enforced by the Pauli blocking factor,
$[1-f_{\nu}(E_3)]$. 
The neutrino tensor $L_{\alpha\beta}$ is given by
\begin{equation}
L^{\alpha\beta}= 8[2k^{\alpha}k^{\beta}+(k\cdot q)g^{\alpha\beta}
-(k^{\alpha}q^{\beta}+q^{\alpha}k^{\beta})\mp i\epsilon^{\alpha\beta\mu\nu}
k_{\mu}q_{\nu}] \,,
\end{equation}
where the incoming four-momentum is $k^\alpha$ and the momentum 
transferred to the medium is $q^\alpha$. 
The minus (plus) sign on the final term applies to neutrino 
(anti-neutrino) scattering.

The medium is characterized by the quark polarization
tensor $\Pi_{\alpha\beta}$.  
In the case of free quarks, each flavor contributes a term of the form
\begin{equation}
\Pi_{\alpha\beta}(q)=-i {\rm Tr}_c \int
\frac{d^4p}{(2\pi)^4} {\rm Tr}~[S_0(p)\Gamma_{\alpha} 
S_0(p+q)\Gamma_{\beta}] \,, 
\label{pi_free}
\end{equation}
where $S_0(p)$ is the free quark propagator at finite chemical potential and
temperature.  
The outer trace is over color and simplifies to a $N_c = 3$ degeneracy.
The inner trace is over spin, and the
$\Gamma_\alpha$ are the neutrino-quark vertex functions which determine
the spin channel.
Specifically, the vector polarization is computed by choosing
$(\Gamma_{\alpha}, \Gamma_{\beta}) = ( \gamma_{\alpha}, \gamma_{\beta} )$.
The axial and mixed vector-axial polarizations are similarly obtained from 
$(\Gamma_{\alpha}, \Gamma_{\beta}) = (\gamma_{\alpha}\gamma_5,
\gamma_{\beta}\gamma_5)$ and $(\Gamma_{\alpha}, \Gamma_{\beta}) =
(\gamma_{\alpha}, \gamma_{\beta}\gamma_5)$, respectively.

The free quark propagators in Eq.~(\ref{pi_free}) are naturally modified in a
superconducting medium.  As first pointed out by Bardeen, Cooper, and Schrieffer
several decades ago, the quasi-particle dispersion relation is modified due to
the presence of a gap in the excitation spectrum.  In calculating these
effects, we will consider the simplified case of QCD with two quark
flavors which obey SU(2)$_L \times$ SU(2)$_R$ flavor symmetry, given that
the light $u$ and $d$ quarks dominate low-energy phenomena.  
Furthermore we
will assume that, through some unspecified effective interactions, quarks pair
in a manner analogous to the BCS mechanism \cite{BCS}.  
The relevant consequences of this are the restoration of
chiral symmetry (hence all quarks are approximately massless) and 
the existence of an energy gap
at zero temperature, $\Delta_0$, with approximate temperature dependence,
\begin{equation}
\Delta(T) = \Delta_0 \sqrt{ 1 - \left(\frac{T}{T_c}\right)^2 }.
\end{equation}
The critical temperature $T_c \simeq 0.57 \Delta_0$ is likewise taken from 
BCS theory; this relation has been shown to hold for perturbative QCD 
\cite{tc} and is thus a reasonable assumption for non-perturbative physics.

Breaking the SU$_c$(3) color group leads to complications not found in
electrodynamics.  In QCD the superconducting gap is equivalent to a diquark
condensate, which can at most involve two of the three fundamental quark
colors.  The condensate must therefore be colored.  Since the scalar diquark
(in the $\bar{\bf 3}$ color representation) appears to always be the most
attractive channel, we consider the anomalous (or Gorkov) propagator
\cite{propagators}
\begin{eqnarray}
F(p)_{a b f g} &=&  
\langle q_{f a}^T(p) C\gamma_5 q_{g b}(-p) \rangle \nonumber\\
&=& -i \epsilon_{a b 3} \epsilon_{fg} 
\Delta \left(\frac{\Lambda^+(p)}{p_o^2 - \xi_p^2} + 
\frac{\Lambda^-(p)}{p_o^2 - \bar{\xi}_p^2}\right) \gamma_5~C\,. 
\label{a_bcs}
\end{eqnarray}
Here, $a,b$ are color indices, $f,g$ are flavor indices, 
$\epsilon_{abc}$ is the usual anti-symmetric tensor and we have conventionally
chosen 3 to be the condensate color.  This propagator is also antisymmetric
in flavor and spin, with $C=-i\gamma_0\gamma_2$ being the charge conjugation
operator.

The color bias of the condensate forces a splitting of the normal quark
propagator into colors transverse and parallel to the diquark.  Quarks of color
3, parallel to the condensate in color space, will be unaffected and propagate
freely, with
\begin{equation}
S_0(p)^{b g}_{a f}= i\delta_a^b \delta^g_f~
\left(\frac{\Lambda^+(p)}{p_o^2 - E_p^2} + 
\frac{\Lambda^-(p)}{p_o^2 - \bar{E}_p^2}\right)
~(p_\mu\gamma^\mu -\mu \gamma_0)\,.
\label{s_0}
\end{equation}
This is written in terms of the particle and anti-particle projection operators
$\Lambda^+(p)$ and $\Lambda^-(p)$ respectively, where $\Lambda^{\pm}(p)=(1 \pm
\gamma_0\vec{\gamma} \cdot \hat{p})/2$.  The excitation energies are simply  
$E_p = |\vec{p}|-\mu$ for quarks and $E_p = |\vec{p}|+\mu$ for anti-quarks.

On the other hand, transverse quark colors 1 and 2 participate in the diquark
and thus their quasi-particle propagators are given as
\begin{equation}
S(p)^{b g}_{a f}= i\delta_a^b \delta^g_f~
\left(\frac{\Lambda^+(p)}{p_o^2 - \xi_p^2} + 
\frac{\Lambda^-(p)}{p_o^2 - \bar{\xi}_p^2}\right)
~(p_\mu\gamma^\mu -\mu \gamma_0)\,.
\label{s_bcs}
\end{equation}
The quasi-particle energy is $\xi_p = \sqrt{(|\vec{p}|-\mu)^2 +
\Delta^2}$, and for the anti-particle $\bar\xi_p =
\sqrt{(|\vec{p}|+\mu)^2 + \Delta^2}$.

\begin{figure}[t]
\centering
{\epsfig
{figure=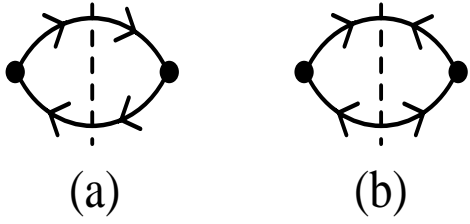,width=0.4\textwidth,height=0.2\textwidth}}
\caption{Standard loop (a) and anomalous loop (b) diagrams contributing to 
the quark polarization operator. }
\label{pol_fig}
\end{figure}

The appearance of an anomalous propagator in the superconducting phase
indicates that the polarization tensor gets contributions from both the normal
quasi-particle propagators (\ref {s_bcs}) and anomalous propagator
(\ref{a_bcs}).  Thus, to order $G_F^2$, Eq.~(\ref{pi_free}) is replaced with
the two contributions corresponding to the 
diagrams shown in Fig.~\ref{pol_fig}, and written
\begin{equation}
\Pi_{\alpha\beta}(q) = -i \!\int\! \frac{d^4p}{(2\pi)^4}  
\left\{
{\rm Tr}~[S_0(p)\Gamma_{\alpha} S_0(p+q)\Gamma_{\beta}] +
2 {\rm Tr}~[S(p)\Gamma_{\alpha} S(p+q)\Gamma_{\beta}] + 
2 {\rm Tr}~[F(p)\Gamma_{\alpha} \bar{F}(p+q)\Gamma_{\beta}]
\right\}
 \,.
\label{pi_bcs}
\end{equation}
The remaining trace is over spin, as the color trace has been performed.
Fig.~\ref{pol_fig}(a) corresponds to the first two terms, which have been 
decomposed into one term with ungapped propagators (\ref{s_0}) and the other
with gapped quasi-particle propagators (\ref{s_bcs}).
Fig.~\ref{pol_fig}(b) represents the third, anomalous term.

For neutrino scattering we must consider vector, axial, and mixed
vector-axial channels, all summed over flavors.  
The full polarization, to be used in evaluating Eq.~(\ref{dcross}),
may be written
\begin{equation}
\Pi_{\alpha\beta} = \sum_f\,\left[ (C_V^f)^2 \Pi^V_{\alpha\beta} +
(C_A^f)^2 \Pi^A_{\alpha\beta} - 2 C_V^f C_A^f\Pi^{VA}_{\alpha\beta}\right]\,.
\label{polsum}
\end{equation}
The coupling constants for up quarks are $C_V^u = \thalf - \fourthr
\sin^2\theta_W$ and $C_A^u = \thalf$ , and for down quarks,
$C_V^d = -\thalf+\twothr\sin^2\theta_W$ and $C_A^d = -\thalf$, 
where $\sin^2\theta_W \simeq 0.23$ is the Weinberg angle.

If we specify a frame in which the transfer momentum is 
$q_{\mu} = (q_0,q,0,0)$ we can separate longitudinal 
components as 
\begin{equation}
\Pi_L^V = - \frac{q_{\mu}^2}{q^2}\Pi^V_{00} = 
- \frac{q_{\mu}^2}{q_0^2}\Pi^V_{11} = 
- \frac{q_{\mu}^2}{q_0q}\Pi^V_{10} \,,
\label{long_def}
\end{equation}
and the transverse,
\begin{equation}
\Pi_T^V =  \Pi_{22}^V = \Pi_{33}^V \,.
\label{trans_def}
\end{equation}
Identical definitions apply to the axial polarizations $\Pi_L^A$ and
$\Pi_T^A$.  The non-zero mixed correlation function is written
\begin{equation}
\Pi_{\alpha\beta}^{VA} = i \epsilon_{\alpha\beta\mu 0}\,q^{\mu}\Pi^{VA}\,.
\label{va_def}
\end{equation}
Detailed calculations of these polarization functions are given in the 
Appendix.  

\begin{figure}[t]
\centering
\leavevmode
{\epsfig
{figure=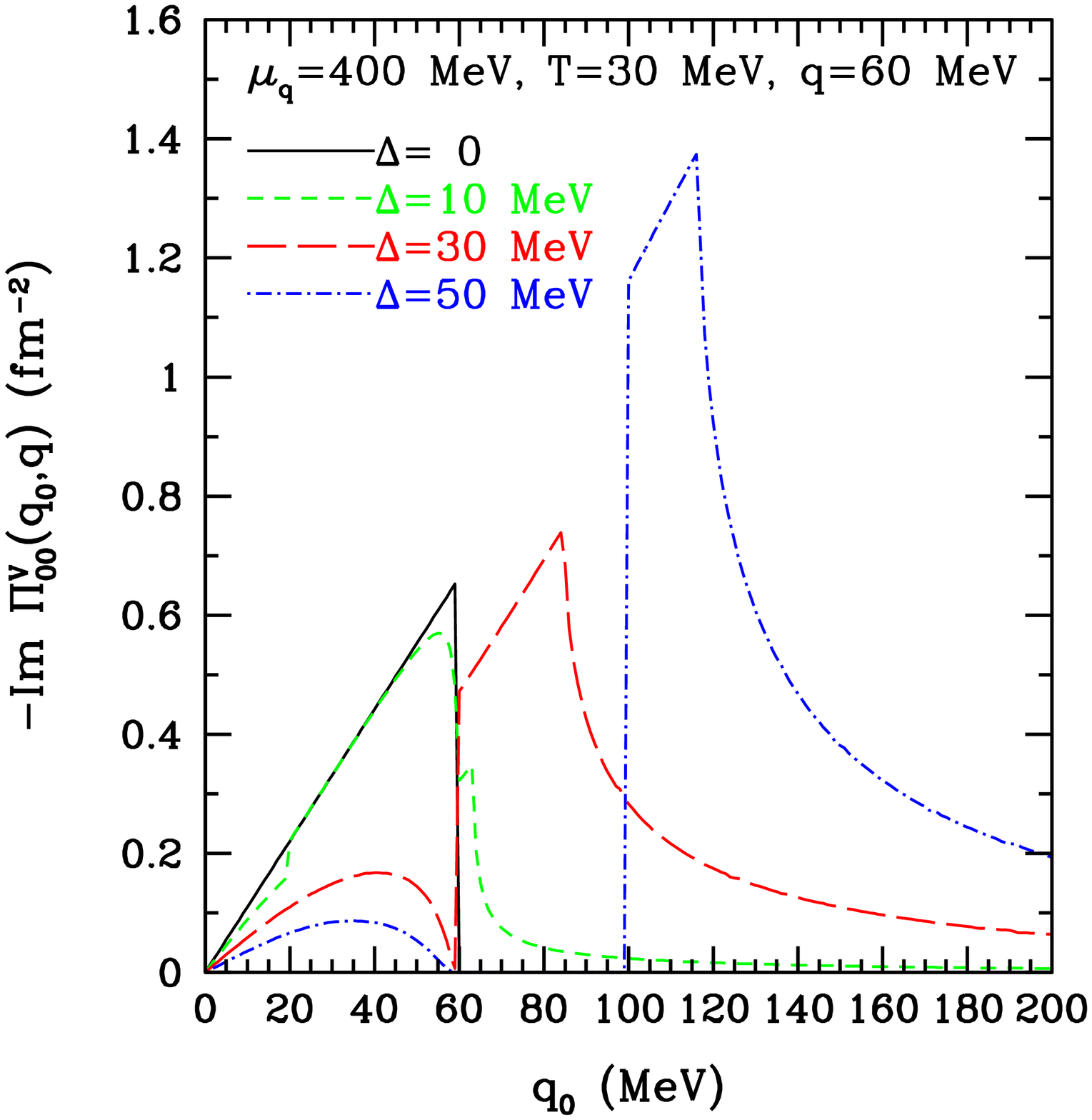,width=0.49\textwidth,height=0.49\textwidth}}
\leavevmode
{\epsfig
{figure=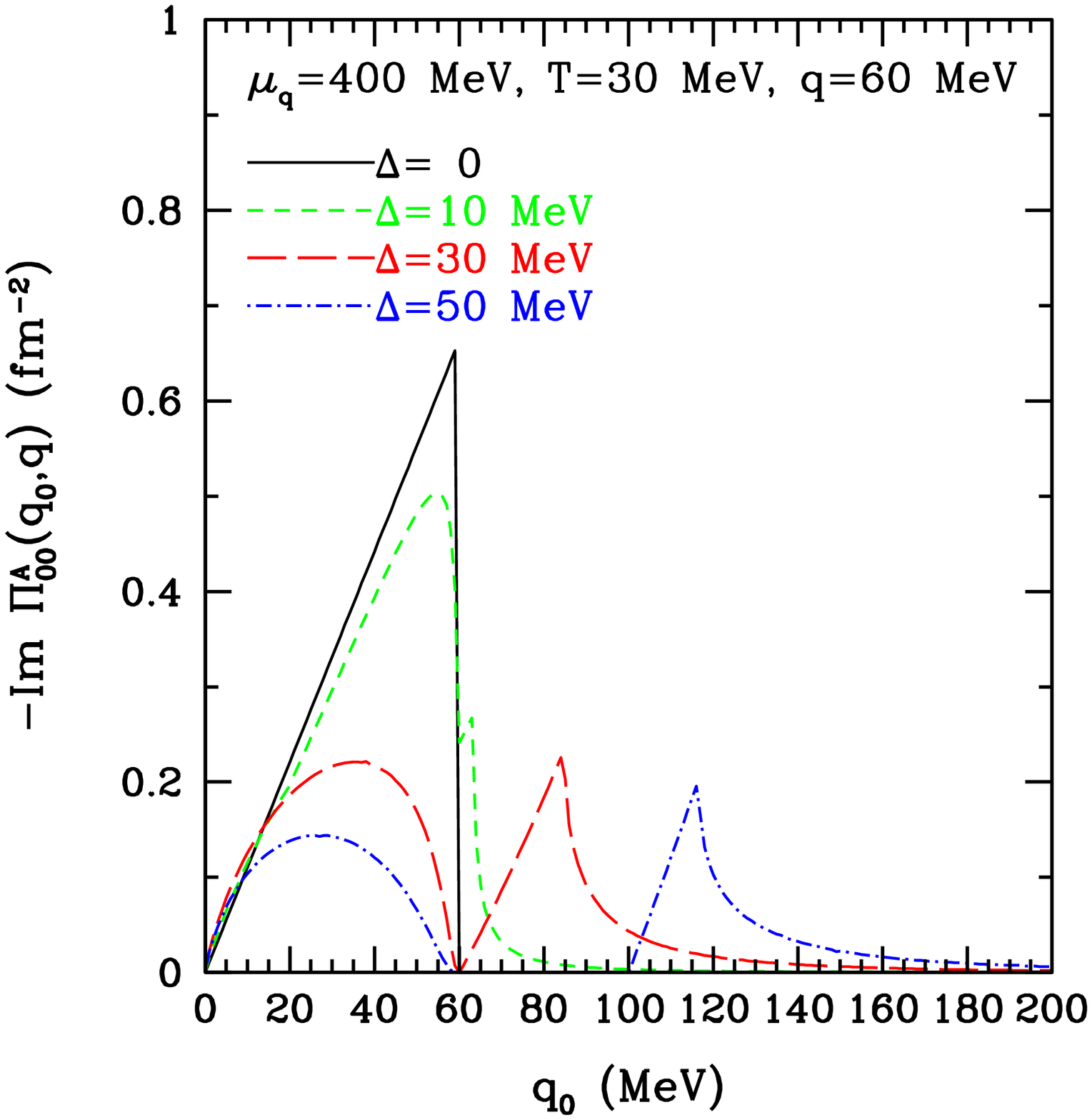,width=0.49\textwidth,height=0.49\textwidth}}
\caption{ Vector-longitudinal (left) and axial-longitudinal (right)
responses as a function of energy transfer. The momentum transfer is
fixed at $q=50$ MeV.}
\label{pizz}
\end{figure}

As the gap increases, the superconducting quasi-particles naturally become the
dominant excitations of the background, a property clearly visible in the
neutrino response functions.  The left panel of Fig.~\ref{pizz} shows the
longitudinal response in the vector channel.  The free quark case, shown as a
solid line labeled $\Delta=0$, is the standard result describing Pauli-blocking
and kinematics of massless single particle excitations \cite{fetwal}.  Here,
energy-momentum conservation restricts the response to the spacelike
region ($q_0<q$).  Superconductivity modifies this result, as the quasi-quark
excitations become suppressed due to the pairing correlations at the Fermi
surface.  At the same time, the response is enhanced when $q_0 \ge 2\Delta$,
signifying the excitation of Cooper pairs.  In particular, the results with
greater $\Delta$ clearly show the threshold for these excitations at energy
transfer $q_0 = 2 \Delta$.  Results for $\Delta=10,\,30$ and $50$ MeV show the
gradual reallocation of response strength from small $q_0$ to the region $q_0
\ge 2\Delta$.  Since scattering probes only the spacelike region, the
$\Pi^V_{00}$ contribution to the cross section is generically {\it suppressed}
in the superconducting phase.  Contributions at $q_0\ge q$ will contribute to
the neutrino production rate, rather than scattering cross section, when the
temperature is near $\Delta$.
Analytic results may be obtained for small transfer energies.
When $q_0$ is much smaller than all other energy scales, the vector 
longitudinal response (see Eq.~(\ref{veclong}) of the Appendix) reduces to
\begin{eqnarray}
\lim_{q_0\rightarrow 0} {\rm Im} ~\Pi^V_L(\Delta) &=&  
\frac{2}{1 + e^{\beta\Delta}}\, \Pi^V_L(\Delta=0) \nonumber\\
&=& \frac{\mu^2 q_0}{2\pi q} \frac{1}{1 + e^{\beta\Delta}}\,.
\end{eqnarray}
From this the weakening of the low energy, vector-longitudinal response
can be calculated for a given gap $\Delta$.

The axial-longitudinal response, which physically corresponds to the 
excitation of spin waves, is shown in the right panel of Fig.~\ref{pizz}.
As with the vector channel a threshold of
$2\Delta$ is apparent but, unlike the previous case, the response as $q_0
\rightarrow 0 $ is {\it enhanced}. 
This is manifest in the the limit $q_0 \ll T \sim \Delta$, where one finds
\cite{lp}
\begin{equation}
{\rm Im} ~\Pi^A_L(\Delta) \simeq
\frac{\Delta}{2T} \,{\rm sech}^2\left(\frac{\Delta}{2T}\right)
{\rm ln}\left(\frac{\Delta}{q_0}\right) \Pi^A_L(\Delta = 0)  \,,
\label{pitqo}
\end{equation}
where in this limit $\Pi_L^A(\Delta=0)=\mu^2 q_0^2 / 4 \pi q$.
This logarithmic enhancement will lead to an integrable peak in the 
differential cross section at $q_0 = 0$, as will be described below.

The transverse response functions, $\Pi^V_T$ and $\Pi^A_T$ as defined
in Eq.~(\ref{trans_def}), exhibit behavior similar to the longitudinal
channels.
The primary distinction is that superconductive coherence at low $q_0$
enhances the vector channel and suppresses the axial.
In the former case the interference effects are constructive and in the 
latter case they are destructive \cite{lowtem3}, which is a reversal of the 
longitudinal results.
This difference has been theoretically understood and experimentally
observed in electric superconductors, where the 
absorption of acoustic waves (vector-longitudinal) is suppressed
while the infra-red absorption (vector-transverse) is enhanced
for small energy transfer \cite{matbar,palmtin}\footnote{
There is naturally a vast literature concerning coherence effects
in superconducting metals, and we find it both gratifying and reassuring
to obtain similar results in this less terrestrial context.}.

In addition to the vector and axial polarizations, the mixed vector-axial 
polarization (\ref{va_def}) also contributes to the cross section.
While this contribution is always much smaller -- by at least one order
of magnitude -- we note that it is enhanced in the superconducting phase,
leading to an amplified difference between neutrino and anti-neutrino
cross sections in PNS matter.

\begin{figure}[ht]
\centering
{\epsfig
{figure=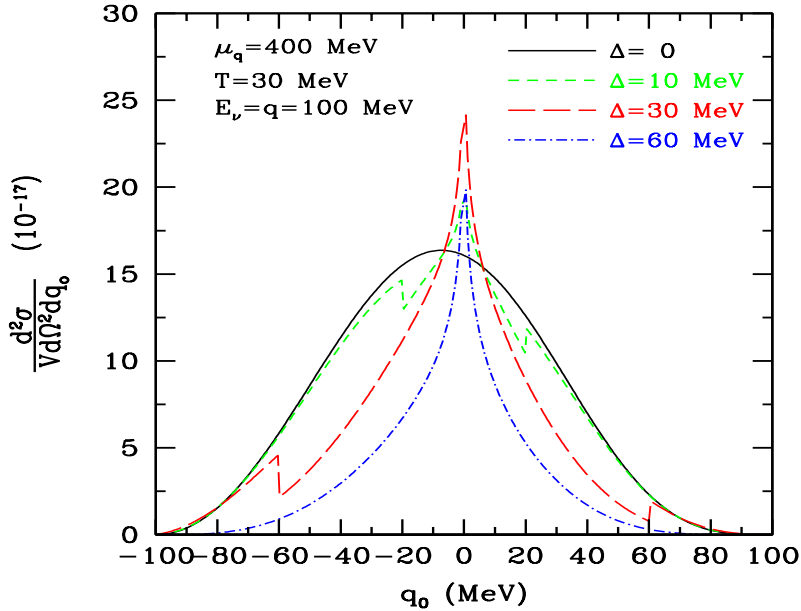,width=0.6\textwidth,height=0.47\textwidth}}
\caption{ Differential cross section as a function of energy transfer and for 
neutrino energy $E_\nu=50$ MeV.}
\label{dsigma}
\end{figure}

Once the polarization tensors have been computed it is straightforward
to obtain the differential cross section, Eq.~(\ref{dcross}). 
Results for neutrinos of energy $E_\nu=50$ MeV,
ambient matter conditions of $\mu_q=400$ MeV and $T=30$ MeV, and 
gaps of varied size are plotted as a function of transfer energy
in Fig.~\ref{dsigma}.
A striking feature of these results is the singular behavior of the
differential cross section near $q_0 = 0$ in superconducting matter.  This is
the (integrable) logarithmic divergence of $\Pi^A_L$ and $\Pi^V_T$ (see
(Eq.~\ref{pitqo})). The threshold behavior seen in differential cross section,
for the case $\Delta=10$ and $30$ MeV, at $q_0=2 \Delta$ and $q_0=-2 \Delta$
correspond to excitations of the Cooper pairs.

The total cross section (per unit volume) is obtained by integrating over all
neutrino energy transfers and angles.  From this the mean free path is
determined, since 
\begin{equation}
\lambda = \left(\frac{\sigma}{V}\right)^{-1}\,.
\end{equation}
As suggested by the
differential cross section in Fig.~\ref{dsigma}, the total cross section is
reduced in the presence of a gap $\Delta$.  The logarithmic peak at $q_0=0$ has
a minimal effect after integration, when $\Delta \sim T$ and when the neutrino
energy $E_\nu \sim \pi T$ since neutrino probe a significant part of the
response outside of this $q_0=0$ region.

\begin{figure}[t]
\centering
\leavevmode
{\epsfig
{figure=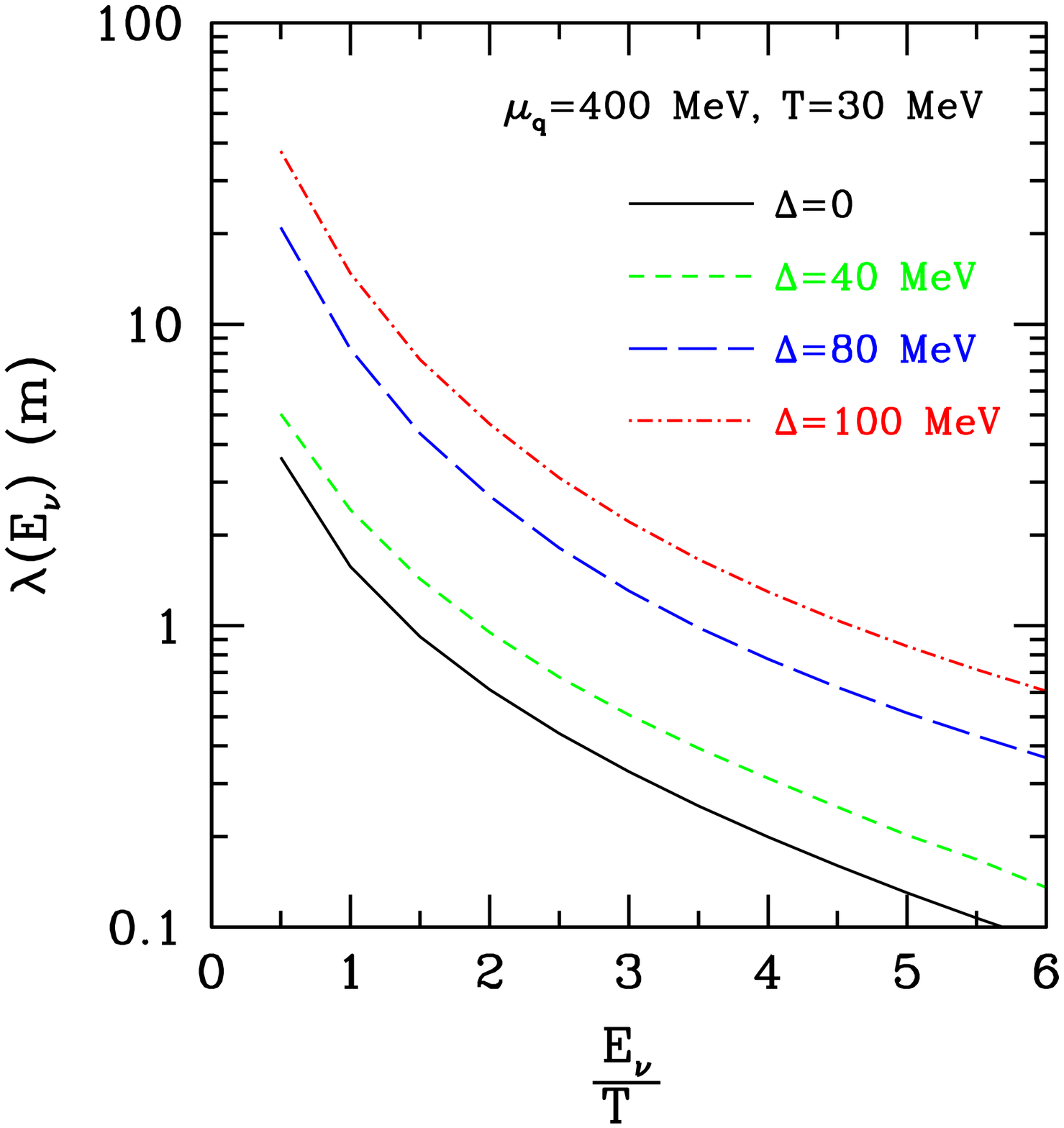,width=0.47\textwidth,height=0.49\textwidth}}
\leavevmode
{\epsfig
{figure=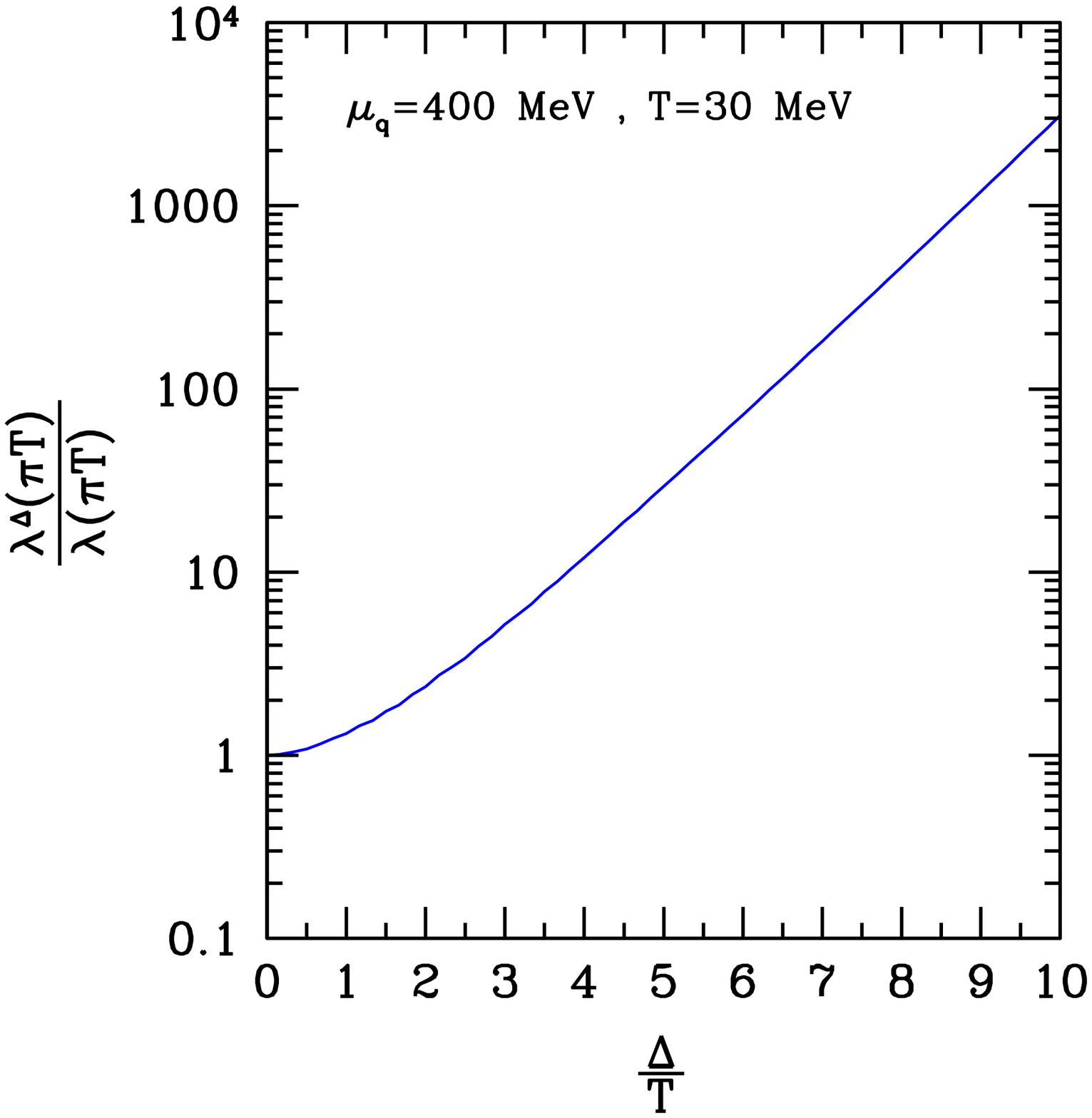,width=0.47\textwidth,height=0.49\textwidth}}
\caption[]
{Left Panel: Neutrino mean free path as a function of neutrino energy. 
Right Panel: The neutrino mean free path for neutrino energy $E_\nu=\pi T$, 
plotted as a function of the gap $\Delta$. 
Results, shown here for $T=30$ MeV, are virtually independent of temperature 
for $T \lsim 50$ MeV.}
\label{lambda}
\end{figure}

Results for the neutrino mean free path, $\lambda$, are shown in
Fig.~\ref{lambda} as a function of incoming neutrino energy $E_\nu$ (ambient
matter conditions of $\mu_q=400$ MeV and $T=30$ MeV have again been used).  The
same energy dependence has been computed previously in the case of free
relativistic and degenerate fermionic matter \cite{rpl}; it decreases as
$1/E_\nu^2$ for $E_\nu \gg T$, and $1/E_\nu$ at $E_\nu \ll T$.  The results
indicate that this energy dependence is not modified by the presence of a gap
when $\Delta \sim T$.  Thus the primary effect of the superconducting phase is
a much larger mean free path.  This is consistent with the suppression found in
the vector-longitudinal response function, which dominates the sum polarization
sum (\ref{polsum}), at $q_0<q$.

\section{The Cooling of an Idealized Quark Star}

The immediate application of the last section's results is neutrino 
emission from a proto-neutron star containing quark matter.
While we will fall far short of a complete description of the role of
color superconductivity in such a complicated environment, in this 
section we will outline the key ingredients for estimating the possible
physical observables.
The first subsection describes a simple and general model for the cooling
of quark matter through temperatures relevant to proto-neutron stars, and
the second addresses the applicability of this scenario to more realistic
situations.

\subsection{The Cooling of Superconducting Quark Matter}

Having determined the neutrino mean free path in a color
superconductor, we now consider the cooling of a macroscopic sphere of
quark matter as it becomes superconducting.
As stated previously, this toy model is motivated by the possibility
that the core of neutron star might contain such matter.
Following our preceding calculations, we will consider the relatively
simple case of two massless flavors with identical chemical potentials.
Furthermore, we will disregard the quarks parallel in color to the
condensate; {\it i.e.} we consider a background comprised exclusively
of quasi-quarks.

The cooling of a spherical system of quark matter from $T \sim T_c \sim 50$ 
MeV is driven by neutrino diffusion, for
the neutrino mean free path is much smaller than the dimensions of system
of astrophysical size, and yet several orders of magnitude larger than
the mean free path of the quarks.
The diffusion equation for energy transport by neutrinos in a spherical
geometry is
\begin{eqnarray}
C_V\, \frac{dT}{dt}=-\frac{1}{r^2} \frac{\partial L_{\nu}}{\partial r} \,,
\label{ediff}
\end{eqnarray}
where $C_V$ is the specific heat per unit volume of quark matter, 
$T$ is the temperature, and $r$ is the radius.
The neutrino energy luminosity for each neutrino type, $L_\nu$,  
depends on the neutrino mean free path and the spatial 
gradients in temperature and is approximated by
\begin{equation}
L_\nu \cong -6\int dE_\nu\,\frac{c}{6\pi^2}\, E_\nu^3 r^2 \lambda(E_\nu)
\frac{\partial f(E_\nu)}{\partial r} \,,
\label{eflux}
\end{equation}
where $c$ denotes the speed of light in vacuum.
In our analysis we assume that neutrino interactions are dominated by the
neutral current scattering common to all neutrino types. 
Consequently, we take the same neutrino and anti-neutrino mean free path 
for every neutrino flavor, giving rise to the factor of six
in Eq.~(\ref{eflux}).  
The equilibrium Fermi distribution, $f(E_\nu)$, and the (scattering) mean
free path, $\lambda(E_\nu)$, are integrated over all neutrino energies,
$E_\nu$.

The solution to the diffusion equation will depend on the size of the
system and its initial temperature gradients.  
However, we are merely interested in a qualitative description of cooling
through a second-order phase transition to superconducting matter.
The temporal behavior is characterized by a time scale $\tau_c$, 
which is proportional to the inverse cooling rate and can hence be deduced
from Eq.~(\ref{ediff}). 
The characteristic time 
\begin{equation}
\tau_c(T)= C_V(T) \frac{R^2}{c\langle\lambda(T)\rangle} \,,
\label{tauc}
\end{equation}
is a strong function of the function of the ambient matter temperature since it
depends on the matter specific heat and the neutrino mean free path.  This
applies to a system characterized by the radial length $R$ and the
energy-weighted average of the mean free path, $\langle\lambda(T)\rangle$.
Following our general treatment of the superconducting gap, we assume that the
temperature dependence of the specific heat is described by BCS theory.  We
will then use the results obtained in the previous section to calculate
$\langle \lambda(T)\rangle$.  Furthermore, since neutrinos are in thermal
equilibrium for the temperatures of interest, we may assume
\begin{equation}
\langle\lambda(T)\rangle \simeq \lambda(E_\nu=\pi T)\,.  
\end{equation}
The quantity on the left is dependent on the gap $\Delta$, a dependence
computed in the previous section and plotted in the right panel of
Fig.~\ref{lambda}.  The results indicate that for small $\Delta/T$ the neutrino
mean free path is not strongly modified, but as $\Delta/T$ increases so too
does $\lambda$, non-linearly at first and then exponentially for $\Delta/T
\gsim 5$.  We note that the diffusion approximation is only valid when $\lambda
\ll R$ and will thus fail for very low temperatures, when $\lambda \sim R$.

\begin{figure}[t]
\centering
{\epsfig
{figure=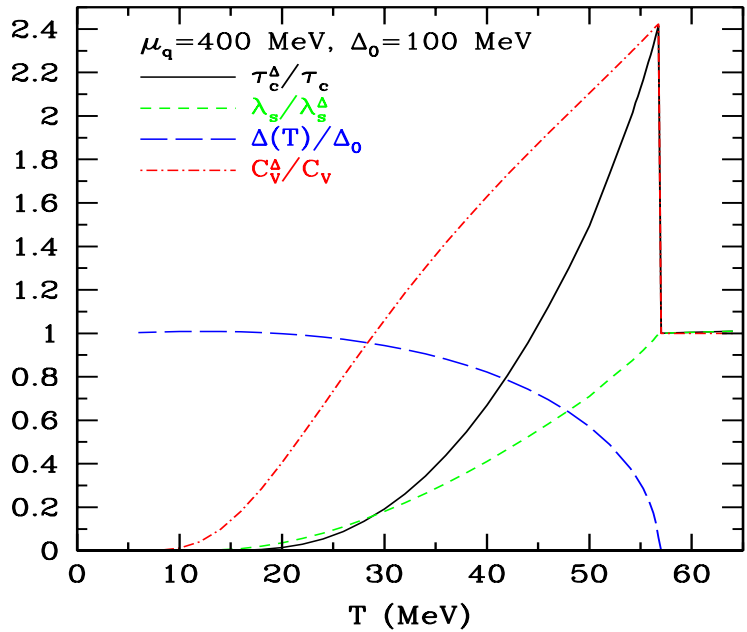,width=0.6\textwidth,height=0.49\textwidth}}
\caption[]{
The figure shows the extent to which different physical quantities are affected
due to the superconducting transition.  Ratios of the cooling time scale (solid
curve), the inverse mean free path (short-dashed curve) and the matter specific
heat (dot-dashed curve) in the superconducting phase to that in the normal
phase is shown as a function of the matter temperature. The ratio of the gap to
it zero temperature value $\Delta_0$ is also shown (long-dashed curve). The
quark chemical potential is $\mu_q=400$ MeV and $\Delta_0=100$ MeV. }
\label{temp}
\end{figure}

The ratio $\tau^{\Delta}_c(T)/\tau_c(T)$, a measure of the extent to
which the cooling rate is changed by a gap, is shown by the solid line in
Fig.~\ref{temp}. 
The ratio $\lambda/\lambda^\Delta$, plotted with the short-dashed curve,
measures the decrease in neutrino interaction rates in the superconducting
background.
The temperature dependences we have taken from BCS theory,
that of the specific heat (dashed curve) and the 
magnitude of the gap itself (dot-dashed curve), are shown for reference in
Fig.~\ref{temp}.
The results shown in Fig.~\ref{temp} are readily interpreted.  The cooling rate
around $T_c$ is influenced mainly by the peak in the specific heat associated
with the second order phase transition, since the neutrino mean free path is
not strongly affected when $\Delta \ll T$.  Subsequently, as the matter cools,
both $C_V$ and $\lambda^{-1}$ decrease in a non-linear fashion for $\Delta \sim
T$.  Upon further cooling, when $\Delta \gg T$, both $C_V$ and $\lambda^{-1}$
decrease exponentially.  Both of these effects accelerate the cooling process.

We conclude that if it were possible to measure the neutrino luminosity from
the hypothetical object described here, a unique temporal profile would be
observed.  This suggests that if a second order, superconducting transition
were to occur inside a proto-neutron star it could be identified by the
temporal characteristics of the late time supernova neutrino signal.
Specifically, there would be a brief interval during which the cooling would
slow around $T\sim T_c$, signified by a period of reduced neutrino detection.

\subsection{Relation to Proto-Neutron Stars}

The temporal pattern of neutrino emission deduced here would be observable 
evidence of the onset of color superconductivity in dense matter.  
However, ours is a simple model, and we must temper our speculations with
considerations of more realistic systems.

Neutrinos emitted from the core of a proto-neutron star, be it hadronic or
quark matter in any phase, must pass through a large amount of matter in the
outer shell.  The neutron-rich material is characterized in the outer shell is
opaque to neutrinos and will thus even out any sharp temporal features
associated with the interior emission.  This is the first and foremost of
caveats since this will directly impact the possibility of observing a
characteristic in the neutrino signal associated with the phase transition.

As discussed in Section II, the scalar diquark condensate in a two-flavor color
superconductor is necessarily colored.  Therefore quarks of one color, taken
conventionally as color 3, will be color-orthogonal to the scalar condensate
and can remain ungapped.  We have not included these color-3 quarks in our
analysis of neutrino scattering since their fate is uncertain \cite{sann}; 
they could form
color-symmetric Cooper pairs, or perhaps a chiral condensate.  If we take the
simplest scenario and assume that they remain free, their presence in the
medium will further dilute any direct effects of the gapped quasi-quarks.
Specifically, the physical quantities in plotted in Fig.~\ref{temp} will not
vanish when $T \rightarrow 0$, instead being reduced to one third of their
$\Delta=0$ values.  Likewise, the maxima at $T_c$ will be reduced relative to a
color-neutral background.

The second-order phase transition is taken directly from BCS theory, which we
assume rather than derive.  While this is the mean-field result for two
massless flavors with equivalent chemical potentials, neutron star matter is
constrained in two ways.  First of all, weak-interaction equilibrium requires
$\mu_d-\mu_u = \mu_e$, where the subscripts refer to down quarks, up quarks,
and electrons.  But since a finite electron number is required to achieve
electric charge neutrality in the star, $\mu_e$ cannot vanish and we
necessarily have $\mu_d\ne\mu_u$.  This difference in chemical potentials is
likely to drive a first order rather than a BCS-second order transition
\cite{paulo}. 

The other notable omission in our study is the strange quark.  As before, weak
interaction equilibrium requires that $\mu_s = \mu_d$, and thus for $\mu_d
\gsim m_s$ strange quarks will be present.  Furthermore, the finite strange
quark mass implies a mismatch in the Fermi momenta which would also drive a
first order transition \cite{strange_quark}.  A generic consequence of a first
order transition is a mixed phase containing both normal and superconducting
quark matter, and transport in the heterogeneous mixed phase is qualitatively
different from that considered here, for neutrino scattering will depend on the
size and nature of the structures (droplets) present.  In previous work it was
shown that neutrino mean free path in the heterogeneous phase can be greatly
reduced due to coherent scattering of droplets in the mixed phase\cite{rbp}.
Combining the results of Section II with neutrino transport in a mixed phase of
superconducting and normal quark matter is beyond the scope of this work.

Therefore, while our calculation of the neutrino mean free path will be an
essential ingredient in a more realistic and hence more complicated model of
neutron star evolution, our toy model only applies to a highly idealized quark
core of a neutron star where $\Delta \gg |\mu_d-\mu_u|$. 

\section{Conclusions}

Motivated by the physical relevance of neutrino diffusion in the cooling of
strongly-interacting matter, we have analyzed the effects on neutrino-quark
scattering arising from a second-order phase transition from normal to color
superconducting quark matter.  The principal microscopic ingredient is the
neutrino mean free path, and this was calculated in linear-response theory with
a BCS-type superconducting background.  We then applied this result to a
schematic model of quark matter at temperatures and densities relevant to the
evolution of proto-neutron stars.  The modified mean free path for neutrinos in
a superconducting background was computed in the simplified approximation of
iso-symmetric, two-flavor quark matter.  We have enumerated the main
shortcomings of these simplifications and realize that a more realistic
treatment of PNS evolution would include many other effects of similar
importance. 

Despite these caveats, our toy model calculation indicates that a
superconducting transition in the quark core of a PNS can impact its early
cooling and thereby potentially alter the temporal characteristics of the
neutrino emission.  We view this work as a first step towards an understanding
of how the presence of color superconducting matter in the core of a neutron
star may affect the early -- and observable -- neutrino signal.  Quark
pairing invariably occurs in theoretical treatments of finite-density QCD, and
thus our microscopic calculations of the neutrino cross sections are pertinent
to transport processes in dense quark matter.  Given the real prospect of
detecting neutrinos emitted from a future supernova event, such transport
processes might someday serve as an probe of the properties of extremely dense
matter.

\section{Acknowledgments}
We thank the organizers of the INT Program on QCD at Finite Baryon Density,
during which this work was begun, and G.W.C. thanks the Institute for Nuclear
Theory for their hospitality. We also thank David Kaplan, M. Prakash, Krishna
Rajagopal, and Martin Savage for critical readings of the manuscript and for
useful comments.  This work was supported by the US Department of Energy grants
DE-FG02-88ER40388 (G.W.C.) and DE-FG03-00-ER41132 (S.R.).

\appendix
\section*{Quark Polarization Tensors}

At densities relevant to this work, only the quark particle-hole excitations
are accessible.  Thus we discard all anti-particle and anti-hole contributions
from the propagators in Eqs.~(\ref{a_bcs}) and (\ref{s_bcs}).

The imaginary part of the vector longitudinal polarization is, for 
each flavor,
\begin{eqnarray}
{\rm Im}~\Pi_{00}^V(q_0,q) &=& 
- 2\pi\int \frac{d^3p}{(2\pi)^3}\frac{d^3k}{(2\pi)^3}
\, \delta^3(p-k-q)(1+\hat{p}\cdot\hat{k}) 
\nonumber\\ &&
\times\Bigg\{ \left[ n(\xi_p) - n(\xi_k) \right] \Bigg[
\delta(q_0+\xi_p-\xi_k)\frac{(\xi_p+E_p)(\xi_k+E_k)-\Delta^2}{4\xi_p\xi_k}
\nonumber\\ && \qquad\qquad
-\delta(q_0-\xi_p+\xi_k)\frac{(\xi_p-E_p)(\xi_k-E_k)-\Delta^2}{4\xi_p\xi_k}
\Bigg]
\nonumber\\ && \qquad
+\left[ 1 - n(\xi_p) - n(\xi_k) \right] \Bigg[
\delta(q_0-\xi_p-\xi_k)\frac{(\xi_p-E_p)(\xi_k+E_k)+\Delta^2}{4\xi_p\xi_k}
\nonumber\\ && \qquad\qquad
-\delta(q_0+\xi_p+\xi_k)\frac{(\xi_p+E_p)(\xi_k-E_k)+\Delta^2}{4\xi_p\xi_k}
\Bigg] \Bigg\} \,.
\label{veclong}
\end{eqnarray}
The axial longitudinal differs only in the sign of the anomalous 
$\Delta^2$ term, a
consequence of coherence effects being different in different channels, and is
given by
\begin{eqnarray}
{\rm Im}~\Pi_{00}^A(q_0,q) &=& 
- 2\pi\int \frac{d^3p}{(2\pi)^3}\frac{d^3k}{(2\pi)^3}
\, \delta^3(p-k-q)(1+\hat{p}\cdot\hat{k}) 
\nonumber\\ &&
\times\Bigg\{ \left[ n(\xi_p) - n(\xi_k) \right] \Bigg[
\delta(q_0+\xi_p-\xi_k)\frac{(\xi_p+E_p)(\xi_k+E_k)+\Delta^2}{4\xi_p\xi_k}
\nonumber\\ && \qquad\qquad
-\delta(q_0-\xi_p+\xi_k)\frac{(\xi_p-E_p)(\xi_k-E_k)+\Delta^2}{4\xi_p\xi_k}
\Bigg]
\nonumber\\ && \qquad
+\left[ 1 - n(\xi_p) - n(\xi_k) \right] \Bigg[
\delta(q_0-\xi_p-\xi_k)\frac{(\xi_p-E_p)(\xi_k+E_k)-\Delta^2}{4\xi_p\xi_k}
\nonumber\\ && \qquad\qquad
-\delta(q_0+\xi_p+\xi_k)\frac{(\xi_p+E_p)(\xi_k-E_k)-\Delta^2}{4\xi_p\xi_k}
\Bigg] \Bigg\} \,.
\label{axlong}
\end{eqnarray}
From these expressions, one may obtain $\Pi_{11}^{V,A}$, $\Pi_{10}^{V,A}$,
and $\Pi_{01}^{V,A}$ as specified in Eq.~(\ref{long_def}).

The transverse response functions have similar forms.
The vector is
\begin{eqnarray}
{\rm Im}~\Pi_{22}^V(q_0,q) &=& 
- 2\pi\int \frac{d^3p}{(2\pi)^3}\frac{d^3k}{(2\pi)^3}
\, \delta^3(p-k-q)(1+\hat{p}\cdot\hat{k} - 2\hat{p}_2\hat{k}_2) 
\nonumber\\ &&
\times\Bigg\{ \left[ n(\xi_p) - n(\xi_k) \right] \Bigg[
\delta(q_0+\xi_p-\xi_k)\frac{(\xi_p+E_p)(\xi_k+E_k)+\Delta^2}{4\xi_p\xi_k}
\nonumber\\ && \qquad\qquad
-\delta(q_0-\xi_p+\xi_k)\frac{(\xi_p-E_p)(\xi_k-E_k)+\Delta^2}{4\xi_p\xi_k}
\Bigg]
\nonumber\\ && \qquad
+\left[ 1 - n(\xi_p) - n(\xi_k) \right] \Bigg[
\delta(q_0-\xi_p-\xi_k)\frac{(\xi_p-E_p)(\xi_k+E_k)-\Delta^2}{4\xi_p\xi_k}
\nonumber\\ && \qquad\qquad
-\delta(q_0+\xi_p+\xi_k)\frac{(\xi_p+E_p)(\xi_k-E_k)-\Delta^2}{4\xi_p\xi_k}
\Bigg] \Bigg\} \,,
\label{vectrans}
\end{eqnarray}
and the axial,
\begin{eqnarray}
{\rm Im}~\Pi_{22}^A(q_0,q) &=& 
- 2\pi\int \frac{d^3p}{(2\pi)^3}\frac{d^3k}{(2\pi)^3}
\, \delta^3(p-k-q)(1+\hat{p}\cdot\hat{k} - 2\hat{p}_2\hat{k}_2) 
\nonumber\\ &&
\times\Bigg\{ \left[ n(\xi_p) - n(\xi_k) \right] \Bigg[
\delta(q_0+\xi_p-\xi_k)\frac{(\xi_p+E_p)(\xi_k+E_k)+\Delta^2}{4\xi_p\xi_k}
\nonumber\\ && \qquad\qquad
-\delta(q_0-\xi_p+\xi_k)\frac{(\xi_p-E_p)(\xi_k-E_k)+\Delta^2}{4\xi_p\xi_k}
\Bigg]
\nonumber\\ && \qquad
+\left[ 1 - n(\xi_p) - n(\xi_k) \right] \Bigg[
\delta(q_0-\xi_p-\xi_k)\frac{(\xi_p-E_p)(\xi_k+E_k)-\Delta^2}{4\xi_p\xi_k}
\nonumber\\ && \qquad\qquad
-\delta(q_0+\xi_p+\xi_k)\frac{(\xi_p+E_p)(\xi_k-E_k)-\Delta^2}{4\xi_p\xi_k}
\Bigg] \Bigg\} \,.
\label{axtrans}
\end{eqnarray}

Finally, there is a small but finite response in the mixed vector-axial
channel.  Since the neutrino tensor element $L^{23}$ is imaginary, we take
the real part of the polarization.
Antisymmetric in spin, it is
\begin{eqnarray}
{\rm Re}~\Pi_{23}^{VA}(q_0,q) &=& 
- 2\pi\int \frac{d^3p}{(2\pi)^3}\frac{d^3k}{(2\pi)^3}
\, \delta^3(p-k-q)(\hat{p}_1 - \hat{k}_1) 
\nonumber\\ &&
\times\Bigg\{ \left[ n(\xi_p) - n(\xi_k) \right] \Bigg[
\delta(q_0+\xi_p-\xi_k)\frac{(\xi_p+E_p)(\xi_k+E_k)+\Delta^2}{4\xi_p\xi_k}
 \nonumber\\ && \qquad\qquad
-\delta(q_0-\xi_p+\xi_k)\frac{(\xi_p-E_p)(\xi_k-E_k)+\Delta^2}{4\xi_p\xi_k}
\Bigg]
\nonumber\\ && \qquad
+\left[ 1 - n(\xi_p) - n(\xi_k) \right] \Bigg[
\delta(q_0-\xi_p-\xi_k)\frac{(\xi_p-E_p)(\xi_k+E_k)-\Delta^2}{4\xi_p\xi_k}
\nonumber\\ && \qquad\qquad
-\delta(q_0+\xi_p+\xi_k)\frac{(\xi_p+E_p)(\xi_k-E_k)-\Delta^2}{4\xi_p\xi_k}
\Bigg] \Bigg\} \,.
\end{eqnarray}

\end{document}